\documentclass[12pt]{iopart}

\usepackage{iopams}
\usepackage{graphicx}
\usepackage{color}
\begin{document}

\title[Statistical analysis of coupled oscillations with a single fixed endpoint]{Statistical analysis of coupled oscillations with a single fixed endpoint and its application to carbon nanomaterials}

\author{Norio Inui}

\address{Graduate School of Engineering, 
University of Hyogo, 
Himeji, Hyogo 671-2280, Japan}
\ead{inui@eng.u-hyogo.ac.jp}
\vspace{10pt}
\begin{indented}
\item[]March 2023
\end{indented}

\begin{abstract}
Squared deviations from the equilibrium positions of 
one-dimensional coupled harmonic oscillators with fixed and 
free endpoints are 
calculated, and the time averages are expressed as a function of the initial displacements and velocities.
Furthermore, we consider the averages of squared deviations over an ensemble of initial displacements and velocities, which distribute based on a product of the same distribution functions with variances of the initial coordinates $\sigma_{x}^2$ and velocities $\sigma_{v}^2$, respectively. 
We demonstrate that the mean squared deviation linearly increases as the oscillator separates further from the fixed endpoint because of the asymmetrical boundary conditions, and 
that the increase rate depends only on $\sigma_{v}^2$ and not on $\sigma_{x}^2$. 
This simple statistical property of  harmonic oscillation is similarly observed in the oscillations of a graphene sheet and  carbon nanotube in molecular dynamics simulations, in which the interacting forces are nonlinear.
\end{abstract}

%
%
%
%
%

\section{Introduction}\label{S1}
Coupled oscillators, in which particles are connected with springs, 
have been used to investigate the statistical properties of various
interacting many-body systems, such as ergodicity \cite{Mazur1960,Poggi1997,Berman2005,Baldovin2023}, transport phenomena \cite{Takizawa1964,Lepri2003,Dhar2008,Novotny2010}, and synchronization \cite{Heagy1994,Shim2007}. The time evolution of a chain of coupled harmonic oscillators is deterministic and 
can be analytically expressed. However, it exhibits complicated behaviours such as  Brownian motion \cite{Wang1945,Spohn1980,Ford2001,Kupfermana2004,Ghosh2023}.
These pseudo-random motions originate from the superposition of harmonic oscillations with irrational frequencies.

The number of oscillation modes
and the degree of initial displacements associated with  coupled oscillators proportionally increase with the particle number. It follows that more complicated temporal changes in the displacement can be generated as the particle number increases.
Thus, statistical analysis is useful for understanding the dynamics of coupled oscillators. 
By calculating the averages over the initial displacement and velocities distributed under a canonical ensemble, Ford, Kac, and Mazur \cite{Ford1965} demonstrated that coupled harmonic oscillators arranged in a chain with cyclic boundary conditions can be used to create a model of a heat bath as the number of oscillators approaches infinity. 

In this study, we investigate the statistical properties of a chain of coupled harmonic oscillators, in which only one of the two endpoints is fixed. In contrast to the case of coupled harmonic oscillators with cyclic boundary conditions, the statistical properties of the displacement from the equilibrium position depend on the distance from the fixed endpoint because of  asymmetrical boundary conditions \cite{Bao2011}. 
We focus on the ensemble average of the squared deviation from the equilibrium position by introducing two averaging procedures:  averaging over time and averaging over the initial coordinates and velocities distributed based on
a specified probability distribution. 

This study demonstrates that the mean squared deviation from the equilibrium position linearly increases  as the position of particle is farther away from the fixed end of the coupled harmonic oscillator. The effectiveness of the application of the coupled harmonic oscillator to actual materials for this behaviour remains to be investigated. Therefore, we examine whether the  mean squared deviation increases linearly, based on molecular dynamics (MD) simulations of the oscillations of graphene sheets and  carbon nanotubes 
\cite{Zou2016, Eftekhari2021, Bachtold2022}.

The remainder of this paper is organized as follows: Section I$\hspace{-0.2mm}$I discusses the equation of motion for one-dimensional coupled harmonic oscillators 
with a fixed endpoint  using a matrix. The displacement from the equilibrium point is expressed as a function of time using the eigenvalues and eigenvectors. Section I$\hspace{-0.2mm}$I$\hspace{-0.2mm}$I presents 
the time average of the squared deviation from the equilibrium position as
a function of the initial displacements and velocities of oscillators of the matrix.
In Section I$\hspace{-0.2mm}$V,
we assume that the  initial displacements of each particle independently obey the same distribution with  standard deviation $\sigma_{x}^2$. Similarly, the initial velocities are assumed to obey the distribution with  standard deviation $\sigma_{v}^2$.
Under these assumptions, we demonstrate that the mean squared deviation from the equilibrium position  increases linearly  for the particle located farther from the fixed endpoint, and that its coefficient can be determined  using only $\sigma_{v}^2$.
In Section V, the mean squared deviation of  the coupled harmonic oscillators, whose mass and spring constants
change alternately, is calculated using Monte Carlo methods and compared with the exact results obtained using a diagonalization method.
In Section V$\hspace{-0.2mm}$I, the oscillations of a graphene sheet and  carbon nanotube are analyzed based on  MD simulations, and the linear increase  is examined to understand the variance in their displacement. Section V$\hspace{-0.2mm}$I$\hspace{-0.2mm}$I summarizes the effects of  asymmetrical boundary conditions on the dynamics of coupled harmonic oscillations and discusses their application to carbon nanomaterials.

\section{Motion of coupled oscillations with a single fixed endpoint}\label{S2}
A one-dimensional coupled harmonic oscillator comprises $N$ particles with mass $m$ and $N$ linear springs, whose equilibrium length is $l_{0}$ and spring constant is $K$, as shown in Fig. 1. 
The left end of the first spring is fixed, and the particle on the far right is 
connected  only to its left  particle. 

\begin{figure}[h]
\hspace{20mm}
\begin{center}
\includegraphics[bb= 0 0 380 50 , origin = c, clip,scale=0.9]{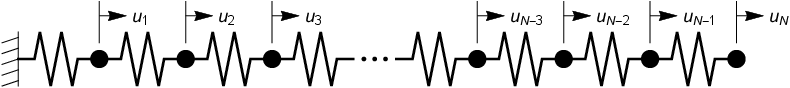}
\end{center}
\caption{Chain of coupled harmonic oscillators comprising $N$ particles, which are represented by sold circles. The displacement of the $n$-th particle is denoted by $u_{n}$. The left end of the first spring is fixed. }
\end{figure}

The displacement of the $n$-th mass at time $\tau$ is denoted as $u_{n}(\tau)$.
The equations of motion for coupled oscillations with a single fixed endpoint are  as follows:

\begin{eqnarray}
m\frac{d^2u_{1}(\tau)}{d\tau^2} &=& -2Ku_{1}(\tau)+Ku_{2}(\tau), \\
m\frac{d^2u_{n}(\tau)}{d\tau^2} &=& Ku_{n-1}(\tau)-2Ku_{n}(\tau)+ Ku_{n+1}(\tau) , 
\hspace{3mm} 2\leq n \leq N-1,\\
m\frac{d^2u_{N}(\tau)}{d\tau^2} &=& Ku_{N-1}(\tau) -Ku_{N}(\tau).
\end{eqnarray}
To express these equations with nondimensional valuables, 
a dimensionless time $t$ $\equiv$ $\tau/\omega_{0}$, 
where $\omega_{0}$ is defined as $\sqrt{K/m}$, and a normalized displacement defined by
$x_{n}(t)$ $\equiv$ $u(\tau)/l_{0}$ are introduced. In addition, the normalized displacement vector is defined as follows:
\begin{eqnarray}
\vec{x}(t) = (x_{1}(t), x_{2}(t), \ldots, x_{N}(t))^{T},
\end{eqnarray}
where $T$ represents the transpose of a vector.
The dimensionless differential equation of $\vec{x}(t)$ is expressed as 
\begin{eqnarray}
\frac{d^2\vec{x}(t)}{dt^2} &=& M\vec{x}(t),
\label{Deq2}
\end{eqnarray}
where $M$ represents an $N \times N$ matrix defined by
\begin{eqnarray}
M &=&
\left [
\begin{array}{cccccccc}
-2 & 1 & 0 & 0 & \cdots & 0 & 0 & 0\\
1 & -2 & 1 & 0 & \cdots & 0 & 0 & 0\\
0 & 1 & -2 & 1 & 0 & \cdots & 0 & 0\\
0 & 0 & 1 & -2 & 1 & 0 & \cdots & 0\\
\vdots & \vdots & \vdots & \vdots & \vdots & \vdots & \vdots & \vdots\\
0 & 0 & 0 & 0 & \cdots & -2 & 1 & 0\\
0 & 0 & 0 & 0 & \cdots & 1 & -2 & 1\\
0 & 0 & 0 & 0 & \cdots & 0 & 1 & -1\\
\end{array}
\right ].
\label{matrix1}
\end{eqnarray}

The solution of Eq. (\ref{Deq2}) is obtained through the diagonalization of $M$ \cite{Fonseca2020}.
The eigenvalue of $M$ is expressed as 
\begin{eqnarray}
\lambda_{k} &=& -2+2\cos\left[\frac{(2k-1)\pi}{2N+1} \right],
\hspace{3mm} 1 \leq k \leq N.
\label{eigenvalues}
\end{eqnarray}
The $j$-th element of the eigenvectors for $\lambda_{k}$ are represented as follows:
\begin{eqnarray}
u_{j,k} &=& -\frac{2}{\sqrt{1+2N}}(-1)^k
\sin\left[\frac{(2k-1)j\pi}{2N+1} \right], \hspace{3mm} 1 \leq j, k \leq N.
\label{eigenvectors}
\end{eqnarray}
The eigenvectors are normalized, and they satisfy the following formulae: 
\begin{eqnarray}
\sum_{j=1}^{N} u^2_{j,k} &=& 1, \label{normA} \\
\sum_{k=1}^{N} u^2_{j,k} &=& 1. \label{normB} 
\end{eqnarray}
If the initial displacement and velocity of the $n$-th particle are $x_{n}(0)$ and $v_{n}(0)$, respectively, then 
the displacement of the $n$-th particle at time $t$ is expressed as follows: 
\begin{eqnarray}
x_{n}(t) &=& x_{n}^{(1)}(t)+x_{n}^{(2)}(t), 
\label{twoparts} \\
x_{n}^{(1)}(t) &=& \sum_{k=1}^{N}
\left(
\sum_{j=1}^{N}u_{j,k}x_{j}(0)
\right)
u_{n,k}\cos\omega_{k}t, \label{x1} \\
x_{n}^{(2)}(t) &=& \sum_{k=1}^{N}
\left(
\sum_{j=1}^{N}u_{j,k}v_{j}(0)
\right)
\frac{u_{n,k}}{\omega_{k}}\sin\omega_{k}t, \label{x2}
\end{eqnarray}
where $\omega_{k}$ = $\sqrt{-\lambda_{k}}$.


\section{Temporal average of displacements}\label{S3}
The time-dependence of a single harmonic oscillator is  simple. However, coupled oscillators exhibit complex behaviours. 
Figure 2 shows an example of the trajectories of coupled oscillators comprising 
10 particles. In this case, the fluctuation of the 10-th particle is greater than 
that of the first particle. To characterize the dependence of the fluctuation of oscillators on $n$, we 
averaged the displacement over time.

\begin{figure}[h]
\hspace{20mm}
\begin{center}
\includegraphics[bb= 0 0 280 180 , origin = c, clip,scale=1.0]{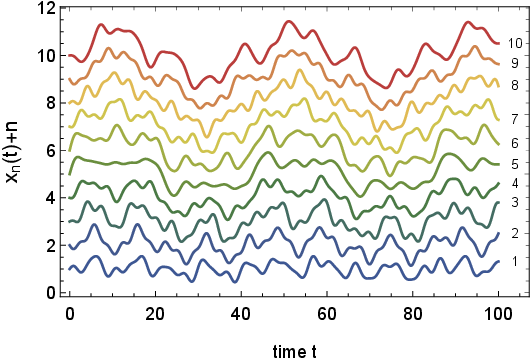}
\end{center}
\caption{(color online) Example of trajectories of coupled harmonic oscillators comprising 10 particles in which the first particle is connected to a fixed endpoint.}
\end{figure}
In statistical physics, the average obtained using a canonical ensemble in an equilibrium state has been discussed \cite{Bao2011}. However, the time evolution of coupled oscillations  depends strongly on the initial conditions. Thus, in this study, we performed time-averaging for fixed the initial conditions, and then considered the ensemble averaging.   This approach allows the time-averaged displacement 
to be easily measured for a macroscopic mass-spring system and then compared with theoretical results.

We define the time average of a time-dependent variable $z(t)$ in  interval $T$ as
\begin{eqnarray}
\langle z \rangle_{T} &\equiv & 
\frac{1}{T}\int_{0}^{T} z(t)dt. 
\label{deftimeaverage}
\end{eqnarray}
Figure 3 shows that the time averaged value of $x^{2}_{n}(t)$, which is the square of $x(t)$ in Fig. 2, converges to a constant value depending on the index of particle $n$ as $T$ increases. We represent the value of $\langle z \rangle_{T}$ in the limit of 
$T \rightarrow \infty$ as $\overline{z}$. 

\begin{figure}[h]
\hspace{20mm}
\begin{center}
\includegraphics[bb= 0 0 280 230 , origin = c, clip,scale=1.0]{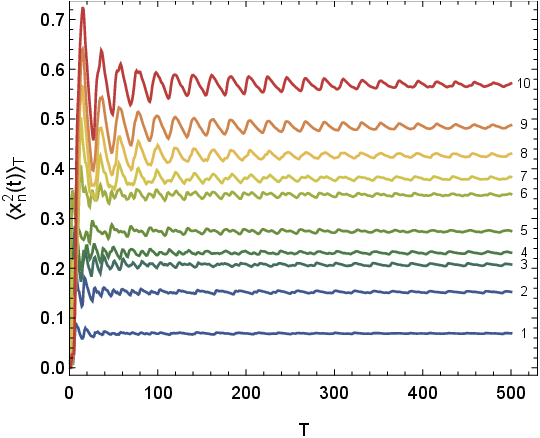}
\end{center}
\caption{(color online) Time average of the squared displacement, 
$\langle x^{2}_{n}(t) \rangle_{T}$ for $x_{n}(t)$ in Fig. 2. }
\end{figure}

The cross-term $x_{n}^{(1)}(t)x_{n}^{(2)}(t)$ exists in the expansion $x_{n}^2(t)$.
However, its contribution to the time-averaged value vanishes via time averaging because $\overline{\sin(\omega_{1}t)\cos(\omega_{2}t)}$ = 0. Thus, $\overline{x_{n}^2}$ 
can be written as the sum of the two contributions:
\begin{eqnarray}
\overline{x_{n}^2} &=& \overline{\left(x_{n}^{(1)}\right)^2}
+\overline{\left(x_{n}^{(2)}\right)^2}.
\end{eqnarray}
Using Eq. (\ref{x1}), the square of $x^{(1)}_{n}(t)$ is written as 
\begin{eqnarray}
(x^{(1)}_{n}(t))^2 &=& 
\left [
\sum_{k_{1}=1}^{N}
\left(
\sum_{j=1}^{N}u_{j,k_{1}}x_{j}(0)
\right)
\cos\omega_{k_{1}}t
\right]
\left [
\sum_{k_{2}=1}^{N}
\left(
\sum_{j=1}^{N}u_{j,k_{2}}x_{j}(0)
\right)
\cos\omega_{k_{2}}t
\right]. \nonumber \\
\end{eqnarray}
If $\omega_{k_{1}}$ $\neq$ 0 or $\omega_{k_{2}}$ $\neq$ 0, the following equation holds:
\begin{eqnarray}
\lim_{T \rightarrow \infty}
\frac{1}{T}\int_{0}^{T}
\cos (\omega_{k_{1}}t)
\cos (\omega_{k_{2}}t)dt
&=&
\left \{
\begin{array}{ll}
\frac{1}{2} & k_{1} = k_{2}, \\
0, & k_{1} \neq k_{2}.
\end{array}
\right. 
\end{eqnarray}
Furthermore, recalling that eigenvalues are given by Eq. (\ref{eigenvalues}), 
we find that $\omega_{k}$ is nonzero for any $k$.
Thus, the time average of $(x^{(1)}_{n})^2$ is given by
\begin{eqnarray}
\overline{\left(x^{(1)}_{n}\right)^2}
&=&
\frac{1}{2}
\sum_{k=1}^{N}
\left(
\sum_{j=1}^{N}u_{j,k}x_{j}(0)
\right)^{2}u_{n,k}^2.
\end{eqnarray}
Similarly, the time average of $(x^{(2)}_{n}(t))^2$ is given by
\begin{eqnarray}
\overline{\left(x^{(2)}_{n}\right)^2}
&=&
\frac{1}{2}
\sum_{k=1}^{N}
\left(
\sum_{j=1}^{N}u_{j,k}v_{j}(0)
\right)^2
\left(
\frac{u_{n,k}}{\omega_{k}}
\right)^2.
\end{eqnarray}


\section{Average over initial displacements and velocities}\label{S4}
The dynamics of coupled oscillators is determined by the initial displacements 
$\vec{X}_{0}$ $\equiv$ 
$\{x_{1}(0),x_{2}(0),$ $\ldots,x_{N}(0)\}$ and  initial velocities
$\vec{V}_{0}$ $\equiv$ $\{v_{1}(0),v_{2}(0),$  $\ldots,v_{N}(0)\}$.
We assume that $\vec{X}_{0}$ and $\vec{V}_{0}$ 
are sets of random variables of a probability distribution 
$P(\vec{X}_{0},\vec{V}_{0})$.
The average of variable $z(\vec{X}_{0},\vec{V}_{0})$ over the initial displacements and velocities is defined as
\begin{eqnarray}
\langle z(\vec{X}_{0},\vec{V}_{0}) \rangle
&\equiv&
\int_{-\infty}^{\infty}
\cdots
\int_{-\infty}^{\infty}
z(\vec{X}_{0},\vec{V}_{0})
P(\vec{X}_{0},\vec{V}_{0})
dx_{1}(0)
\nonumber \\
&&\ldots dx_{N}(0)
dv_{1}(0)\ldots dv_{N}(0). 
\end{eqnarray}
We assume that the probability distribution satisfies  $\langle x_{n}(0) \rangle$ = 0 
and $\langle v_{n}(0) \rangle$ = 0 for any value of $n$.
In addition, we assume that 
the second moments of the initial displacement and  velocity are given by
\begin{eqnarray}
\langle x_{n}(0)x_{n^{\prime}}(0) \rangle
&=&
\left \{
\begin{array}{ll}
\sigma_{x}^2 & n = n^{\prime}, \\
0, & n \neq n^{\prime},
\end{array}
\right. \\
\langle v_{n}(0)v_{n^{\prime}}(0) \rangle
&=&
\left \{
\begin{array}{ll}
\sigma_{v}^2 & n = n^{\prime}, \\
0, & n \neq n^{\prime}.
\end{array}
\right. 
\end{eqnarray}
Using these properties, we have
\begin{eqnarray}
\left\langle
\left(
\sum_{j=1}^{N}u_{j,k}v_{j}(0)
\right)^2
\right\rangle
&=& \sigma_{x}^2 \sum_{j=1}^{N}u^2_{j,k}, \\
&=& \sigma_{x}^2.
\end{eqnarray}
Equation (\ref{normA}) is used  to derive the final equation.
Consequently, the average of $\overline{(x^{(1)}_{n})^2}$ over the initial displacements is given by
\begin{eqnarray}
\left \langle \overline{\left(x^{(1)}_{n}\right)^2} \right \rangle
&=&
\frac{\sigma_{x}^2}{2}
\sum_{k=1}^{N}
u^2_{n,k}, \\
&=& \frac{\sigma_{x}^2}{2}.
\end{eqnarray}
Notably, the averaged displacement is independent of $n$.

The average of $\overline{(x^{(2)}_{n}(t))^2}$ over the initial displacements is given by
\begin{eqnarray}
\left \langle \overline{\left(x^{(2)}_{n} \right)^2} \right \rangle
&=&
\frac{\sigma_{v}^2}{2}
\sum_{k=1}^{N}
\left(
\frac{u_{n,k}}{\omega_{k}}
\right)^2.
\end{eqnarray}
Using Eqs. (\ref{eigenvalues}) and (\ref{eigenvectors}), 
we have
\begin{eqnarray}
\left \langle \overline{\left(x^{(2)}_{n}\right)^2} \right \rangle
&=&
\frac{\sigma_{v}^2}{2N+1}
\sum_{k=1}^{N} \frac{\sin^2\left(\frac{(2k-1)n\pi}{2N+1}\right)}
{1-\cos\left[\frac{(2k-1)\pi}{2N+1} \right]}.
\label{Aveg2A}
\end{eqnarray}
The summation in Eq. (\ref{Aveg2A}) is written as
\begin{eqnarray}
\sum_{k=1}^{N} \frac{\sin^2\left(\frac{(2k-1)n\pi}{2N+1}\right)}
{1-\cos\left[\frac{(2k-1)\pi}{2N+1} \right]} 
&=& \frac{n(2N+1)}{2}.
\label{EqAppendix1}
\end{eqnarray}
The derivation of this equation is explained in Appendix A. Consequently, we obtain
\begin{eqnarray}
\left \langle \overline{\left(x^{(2)}_{n}\right)^2} \right \rangle
&=&
\frac{\sigma_{v}^2}{2}n.
\label{Aveg2B}
\end{eqnarray}
In contrast to the first contribution involving $x^{(1)}_{n}$ to the average, the second contribution depends on the particle index. Statistically, the farther away the particle is from the fixed end, the greater is the generated displacement. By combining the two contributions to the displacement, the average displacement of the $n$-th particle is expressed as follows:
\begin{eqnarray}
\left \langle \overline{x^2_{n}} \right \rangle
&=&
\frac{\sigma_{x}^2}{2}
+\frac{\sigma_{v}^2}{2}n.
\label{Avegetotal}
\end{eqnarray}
Because $\left \langle \overline{x_{n}} \right \rangle$ = 0,
we henceforth denote $\left \langle \overline{x^2_{n}} \right \rangle$  
as the variance

The temporal average of the squared deviation of the summation 
$x_{1}(t)+\ldots+x_{n}(t)$ for $\sigma_{x}$ = $0$ can  be exactly calculated as follows:
\begin{eqnarray}
\overline{
\left(\sum_{l=1}^{n}x_{l}(t)
\right)^2
} &=& \frac{\sigma_{v}^2}{2}
\sum_{k=1}^{N}
\left(
\sum_{l=1}^{n}
\frac{u_{l,k}}{\omega_{k}}
\right)^2, \\
&=& \frac{\sigma_{v}^2}{2N+1}
\left[
\sum_{k=1}^{N}
\sum_{l=1}^{n} 
\frac{
\sin^2\left(\frac{(2k-1)l\pi}{2N+1}\right)
}
{1-\cos\left[\frac{(2k-1)\pi}{2N+1} \right]}
+ \right. \nonumber \\
&&
\left.
2\sum_{k=1}^{N}
\sum_{l=1}^{n} 
\sum_{j=1}^{n-l}
\frac{
\sin\left(\frac{(l+j)(2k-1)\pi}{2N+1}\right)
\sin\left(\frac{l(2k-1)\pi}{2N+1}\right)
}
{1-\cos\left[\frac{(2k-1)\pi}{2N+1} \right]}
\right].
\end{eqnarray} 
Combining Eq. (\ref{EqAppendix1}) and the following equation with
integers $l_{1}$ $\geq$ $l_{2}$ (see Appendix A for details)
\begin{eqnarray}
\sum_{k=1}^{N} 
\frac{
\sin\left(\frac{(2k-1)l_{1}\pi}{2N+1}\right)
\sin\left(\frac{(2k-1)l_{2}\pi}{2N+1}\right)
}
{1-\cos\left[\frac{(2k-1)\pi}{2N+1} \right]} 
&=& \left(\frac{1}{2}+N\right)l_{2},
\label{EqAppendix2}
\end{eqnarray}
we obtain
\begin{eqnarray}
\overline{
\left(\sum_{l=1}^{n}x_{l}(t)
\right)^2
} &=& \frac{\sigma_{v}^2}{12}n(n+1)(2n+1).
\end{eqnarray}


\section{Coupled oscillation with two atomic species}\label{S5}
This section presents two objectives.
First, we  determine whether the linear increase in the variance shown in Eq. (\ref{Avegetotal}) is similarly observed  in the coupled oscillators of  two elements with alternating mass and spring constants \cite{Gonzalez2023}.
Second, we  introduce the Monte Carlo method to calculate the variance in carbon nanomaterials.
The interaction energy between carbon atoms includes angular and torsion terms in addition to the harmonic term. Expressing the motion of carbon nanomaterials  analytically is difficult; hence,  we performed MD simulations and used Monte Carlo methods to calculate the  variance.

We confirm that the linear increase shown in Eq. (\ref{Avegetotal}) can be obtained  using the Monte Carlo method for a chain of coupled harmonic oscillations with  the same mass and spring constants. 
Figure 4(a) shows the variance of the coupled oscillators comprising 10 particles, whose  initial velocities obeyed the standard normal distribution.
Time averaging with $T$ = $10^{4}$ (see Eq. (\ref{deftimeaverage})) and averaging with $10^{3}$ initial velocities in the phase space were applied. The variance of the coordinate $\sigma_{x}$ was set to zero. The solid line in Fig. 4(a) shows the exact position dependence, $n/2$, which is consistent the Monte Carlo simulation results.

\begin{figure}[h]
\hspace{20mm}
\begin{center}
\includegraphics[bb= 0 0 250 470, origin = c, clip,scale=0.9]{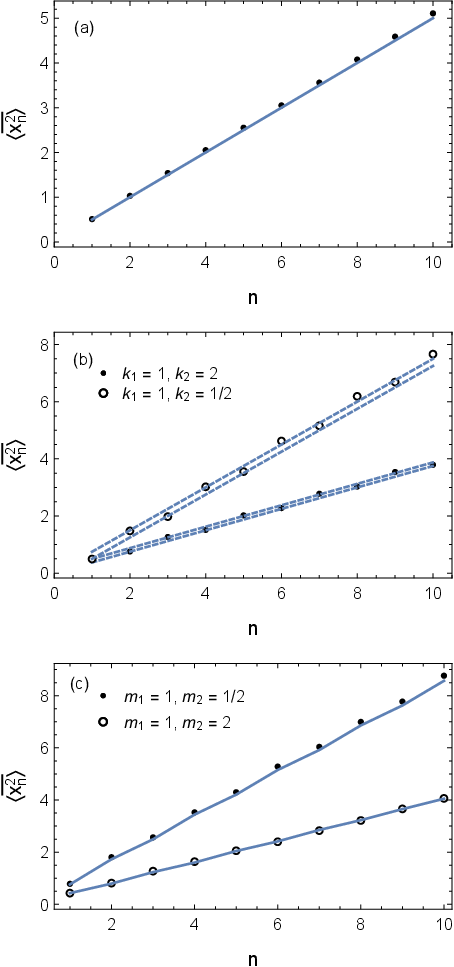}
\end{center}
\caption{(color online) Time variance of harmonic oscillators with (a)  same mass and spring constant, (b) same mass and alternating spring constants, and (c) alternating spring constants and  same mass.}
\end{figure}

Subsequently, we consider a chain of oscillators with alternating spring constant. 
Let $k_{1}$ ($k_{2}$) be the spring constant between
 the $n$-th  and $(n+1)$-th oscillators with even (odd) values of $n$, respectively. If $N$ is even, the matrix describing the dynamics defined in Eq. (\ref{matrix1}) can be  generalized as follows:
\begin{eqnarray}
A &=&
\left [
\begin{array}{cccccccc}
-k_{3} & k_{2} & 0 & 0 & \cdots & 0 & 0 & 0\\
k_{2} & -k_{3} & k_{1} & 0 & \cdots & 0 & 0 & 0\\
0 & k_{1} & -k_{3} & k_{2} & 0 & \cdots & 0 & 0\\
0 & 0 & k_{2} & -k_{3} & k_{1} & 0 & \cdots & 0\\
\vdots & \vdots & \vdots & \vdots & \vdots & \vdots & \vdots & \vdots\\
0 & 0 & 0 & 0 & \cdots & -k_{3} & k_{1} & 0\\
0 & 0 & 0 & 0 & \cdots & k_{1} & -k_{3}& k_{2} \\
0 & 0 & 0 & 0 & \cdots & 0 & k_{2} & -k_{2} \\
\end{array}
\right ],
\label{matrix2}
\end{eqnarray}
where $k_{3}$ = $k_{1}$+$k_{2}$. 
The open and solid circles in Fig. 4(b) show the variance obtained by solving $\ddot{\vec{x}}(t)$ = $A\vec{x}(t)$ and using the Monte Carlo method for $k_{2}$= 1/2 and 2, respectively. The value of parameter $k_{1}$ is 1 for both cases. 
The variances of the oscillators depend on the parity of $n$ but 
the linearity is satisfied for each parity. Matrix $A$ is symmetrical  and diagonalizable using an orthogonal matrix.  
The dashed lines in Fig. 4(b) indicate the exact linear increase in  the set of particles with even and odd values of $n$, separately, which are obtained via a method similar  to that presented in Section V. Their slopes are 3/4 and 3/8 for $k_{2}$= 1/2 and 2, respectively. The slope of  the variances  is independent of the parity of $n$ for each $k_{2}$.

Next, we consider a chain of oscillators with alternating masses. 
Let $m_{1}$ and $m_{2}$ be the masses of particles with even and odd $n$ values, respectively. The matrix element  describing  this system, $B$, is $A_{i,j}/m_{1}$ and $A_{i,j}/m_{2}$ for odd and even $i$ values, respectively, where $A_{i,j}$ is the element of the $i$-th row and $j$-th column of $A$ defined by Eq. (\ref{matrix1}). 
Let $s_{j}$ be an eigenvector of $B$ belonging to eigenvalue $\lambda_{j}$. In contrast to
the eigenvectors of matrices $M$ and $A$,  the orthogonal vectors between different eigenvectors are not satisfied.  However, using  block matrix $S$ = $[s_{1},s_{2},\ldots,s_{N}]$, 
the diagonal matrix can be obtained by calculated from the $S^{-1}BS$.

The solid and open circles in Fig. 4(b) represent the variances obtained using  the Monte Carlo method for 
$m_{1}$= 1/2 and 2, respectively. 
The solid lines in Fig. 4(b), which are obtained by connecting the exact variances obtained through diagonalization are consistent with the results obtained using the Monte Carlo method. Although a parity dependence on $n$ is observed, the variances increases  almost linearly  for each  parity. 

The advantage of the Monte Carlo simulation over calculations using diagonalization is that the variance  distribution of the displacement with respect to time, $\overline{x_{n}}$ can be obtained easily. Figure 5 shows the relative frequency of $\overline{x_{n}}$ for 
a coupled oscillator with a uniform mass and spring constant. The distribution is asymmetric and 
long tailed for large $\overline{x_{n}}$ as $n$ increases.

\begin{figure}[h]
\hspace{20mm}
\begin{center}
\includegraphics[bb= 0 0 450 420, origin = c, clip,scale=0.8]{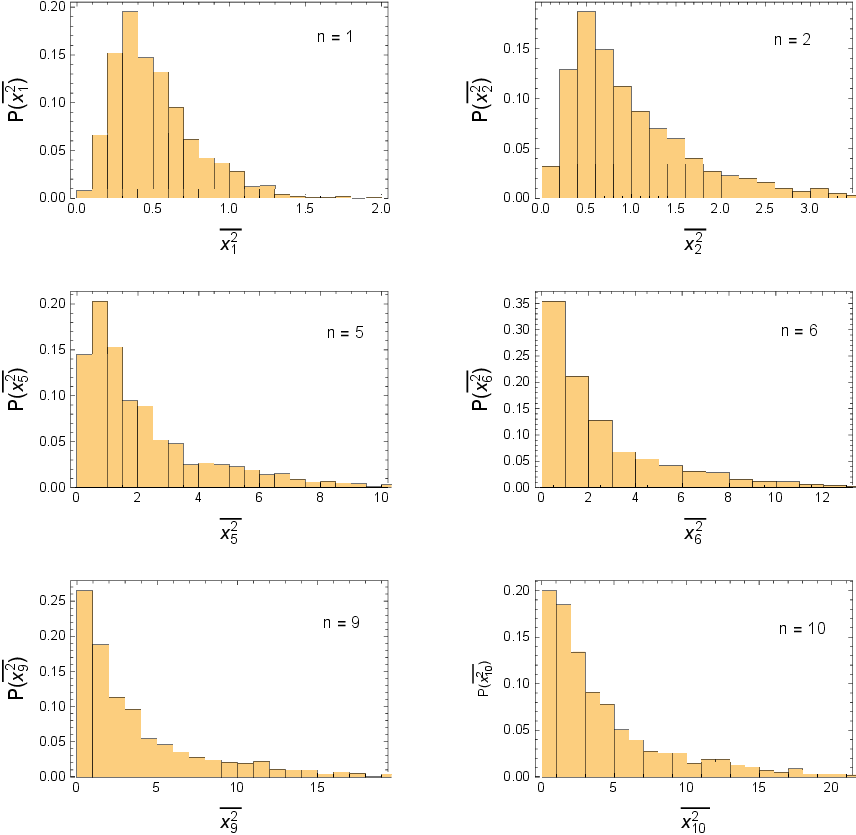}
\end{center}
\caption{(color online) Relative frequency of the time average of 
$\overline{x_{n}^{2}}$ of a harmonic oscillator with same mass and spring constant obtained by the Monte Carlo simulations.}
\end{figure}
We showed that the variance of displacement of the asymmetrical coupled oscillator increased linearly  except when $\sigma_{v}$ $=$ 0.  Under $\sigma_{x}$ = 0, the variances of the velocities are expressed as $\sigma_{v}^2/2$, which is  independent of the position of the oscillator. 


\section{Oscillation of carbon nanomaterials}\label{S6}
In the case of one-dimensional  coupled harmonic oscillations, the variance increased linearly as the oscillator separated from the fixed endpoint when $\sigma_{v}$ $\neq$ 0.
We investigate whether this behaviour is observed  in low-dimensional carbon nanomaterials such as graphene sheets and  
carbon nanotubes.

The configuration of carbon atoms in the graphene sheet considered in this study is shown in Fig. 6.
\begin{figure}[h]
\hspace{20mm}
\begin{center}
\includegraphics[bb= 0 0 210 200, origin = c, clip,scale=0.9]{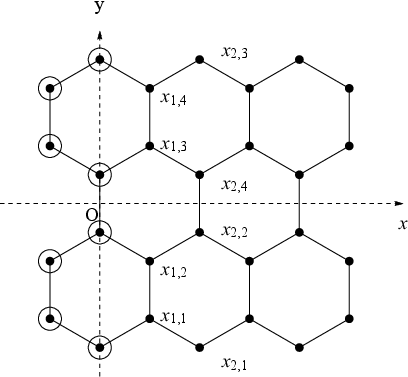}
\end{center}
\caption{Configuration of carbon atoms of a single graphene sheet with 
$N_{x}$ = 5 and $N_{y}$ = 4.
The circles indicate the fixed carbon atoms.}
\end{figure}
The carbon atoms, shown as circles in Fig. 6, are fixed.
The carbon atoms in the equilibrium state are located in a hexagonal lattice, in which 
the $x$- and $y$-axes are set parallel to the zigzag  and armchair edges, respectively, 
and the $z$-axis is set  perpendicular to the sheet.
A set of carbon atoms on a line parallel to the $y$-axis are labelled as $x_{n,1}$, $\ldots$, $x_{n,N_{y}}$, where $N_{y}$ represents the number of  atoms on the $y$-axis for $1$ $\leq$ $n$ $\leq$ $N_{x}$. In addition, the variance of a set of these particles over  time is defined as
\begin{eqnarray}
\overline{X_{n}^2} =\frac{1}{T}\int_{0}^{T} 
\left[
\frac{1}{N_{y}}\sum_{j=1}^{N_{y}}\left(x_{n,j}(t)-\overline{x}_{n,j} \right)^2 
\right]dt.
\label{DefvarG}
\end{eqnarray}
MD simulations were performed to calculate the displacement of the carbon atoms, where the potential between atoms is described by 
the Tersoff potential, which is often used for carbon materials \cite{Tersoff1986,Tersoff1989,Rajasekaran2016}. The force between two carbon atoms depends  on the distance between them and  the bond angles between sets of 
three atoms. Thus, the coupled harmonic oscillators considered in this study are insufficient for accurately describe the graphene sheet. However, if the displacement is small  and restricted to a one-dimensional direction, then the graphene sheet can exhibit  statistical properties similar to those of coupled harmonic oscillators \cite{Bissell2021}. 

We calculate the average of $\overline{X_{n}^2}$ 
for the different initial velocities via MD simulations (micro-canonical ensemble), in which  the time step for numerical integration is set to 0.5 fs and $T$ = 500 ps. The carbon atoms in graphene are strongly combined \cite{Lee2008}, and the fundamental frequency of oscillation in a plane is large \cite{Jiang2009,Chen2013}; therefore, the carbon atoms oscillate repeatedly in 500 ps. To restrict the movement to a one-dimensional direction, 
the initial velocities parallel to the $y$- and $z$-axes
are set to zero. Accordingly, the variance of the initial velocities $\sigma_{y}^2$ and 
$\sigma_{z}^2$ along the $y$- and $z$-axes, respectively are zero. 
In addition, we assume that the same $\sigma_{x}^2$  for all carbon atoms with the same $n$. This restriction is removed in later discussion herein.
The carbon atoms remains at an equilibrium position, i.e.,  $\sigma_{x}$ = 0, for any carbon atoms at $t$ = 0. 

Figure 7 shows the variance over $10^{2}$ sets of initial velocities for graphene sheets with $N_{x}$ = 9 and $N_{x}$ = 6, in which 66 carbon atoms are included. The linear fittings obtained through the least-squares method (solid lines) indicate that the variance along $x$-axis  linearly increase in a wide range of $\sigma_{v,x}$. 
\begin{figure}[h]
\hspace{20mm}
\begin{center}
\includegraphics[bb= 0 0 300 170, origin = c, clip,scale=1.0]{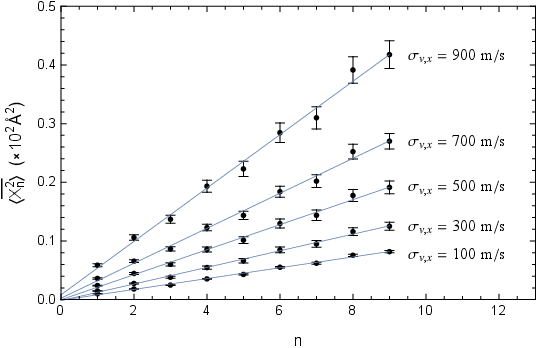}
\end{center}
\caption{(color online) Variance of a graphene sheet obtained through Monte Carlo simulations (solid circles) and linear fittings obtained using the least squares method (solid lines) for different standard deviations of initial velocities.  Error bars show the standard error.}
\end{figure}
Based on  Eq. (\ref{Avegetotal}), the slope of the linear fitting in Fig. 7, $a$,
is expected to be proportional to $\sigma_{v,x}^2$.
Figure 8 depicts the plot of root square of $a$ versus $\sigma_{v,x}$. As expected, $\sqrt{a}$ is proportional to $\sigma_{v,x}$.
\begin{figure}[h]
\hspace{20mm}
\begin{center}
\includegraphics[bb= 0 0 300 165, origin = c, clip,scale=0.9]{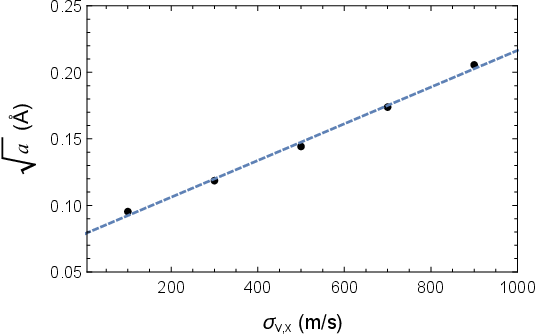}
\end{center}
\caption{(color online) Dependence of the square root of the slope of the linear fittings shown in Fig. 7 on the 
standard deviation of the initial velocity.
The dashed line linear fittings obtained through the least squares method.
}
\end{figure}

The  oscillation  discussed above is one-dimensional.
To consider the effect of two-dimensional motion on the variance,
the initial velocity of each carbon atom is independently chosen from a two-dimensional normal distribution, with $\sigma_{v,x}$ = $\sigma_{v,y}$ =1 km/s for other carbon atoms. The standard deviation $\sigma_{v,z}$ remains zero. Figure 9(a) shows that an almost linear increase occurs even under a two-dimensional motion.
However, the linear increase in the variance of the distance from the $y-$axis, $\langle \overline{r_{n}^{2}} \rangle$   accelerate after adding initial velocities vertical to the surface ($\sigma_{v,z}$ =1 km/s) as shown in 
Fig. 9(b). This is likely caused by the vertical oscillation. The  bending energy is considerably lower than the stretching energy \cite{Lu2009,Wei2013}. Therefore, lateral oscillations can significantly affect the variance. 
\begin{figure}[h]
\hspace{20mm}
\begin{center}
\includegraphics[bb= 0 0 270 310, origin = c, clip,scale=0.9]{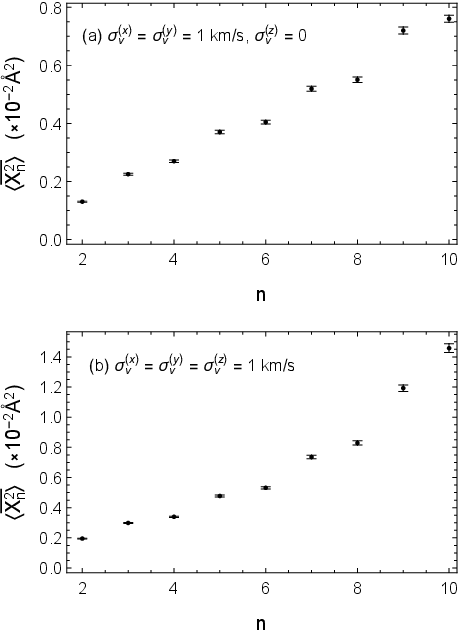}
\end{center}
\caption{Variances of a graphene sheet obtained through Monte Carlo simulations with standard deviations of the initial velocity with 
(a) $\sigma_{v,x}$ = $\sigma_{v,y}$ = 1 km/s and $\sigma_{v,z}$ = 0 and
(b) $\sigma_{v,x}$ = $\sigma_{v,y}$ = $\sigma_{v,z}$ = 1 km/s. Error bars show the standard error.
}
\end{figure}

Additionally, the variance of a carbon nanotube, which is a pseudo-one-dimensional carbon nanomaterial, is calculated. Figure 10 shows the variance parallel to the central axis of a carbon nanotube with zigzag edges(15,0) \cite{Yang2020}. 
The inset of Fig. 10 illustrates the configuration of the carbon atoms, where the solid points indicate  fixed atoms. The three components of the initial velocity  $v_{x}$, $v_{y}$, and $v_{z}$, of each carbon atoms are independently chosen from the normal distribution with the same standard deviation $\sigma_{v}$.
The  black circles in Fig. 10 represent the variances for $\sigma_{v}$ = 100 m/s, and
their increase is slower than the linear increase for large $n$. However,  the variances with large $\sigma_{v}$ ($\sigma_{v}$ = 1 km/s; open circles) exhibits an almost linear increase.
\begin{figure}[h]
\hspace{20mm}
\begin{center}
\includegraphics[bb= 0 0 300 170, origin = c, clip,scale=1.0]{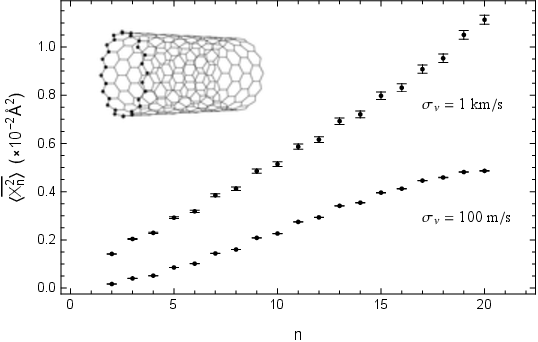}
\end{center}
\caption{Variances of a carbon nanotube obtained through Monte Carlo simulations with standard deviations of the initial velocity with 
(a) $\sigma_{v}$ = 100 m/s and 1 km/s, where all components are randomly chosen from normal distributions. Error bars show the standard error.
}
\end{figure}

\section{Conclusion}
The movement of particles in a coupled oscillator varies depending on their initial displacements and velocity. The motion typically becomes more complicated as the number of particles increases. Therefore, averaging is performed to grasp the statistical properties of the motion. We demonstrated that the average variance of the particles  linearly increases as they are further from the fixed point.
Accordingly, the averaged standard deviation of the $n$-th particle from the fixed end is proportional to $\sqrt{n}$. If the averaged standard deviation is regarded as the characteristic magnitude of the displacement, then the displacement of a particle near the free endpoint increases with chain length. 

Graphene sheets and a carbon nanotube were examined for their linear increase in terms of the averaged variance.
MD simulations yielded similar results as those for a coupled harmonic  oscillator for carbon nanotubes, with a large standard deviation for the initial velocities, which correspond to high temperatures. For the graphene sheet, a linear 
increase of its variance  was observed only when the oscillation is limited in a plane. 
This is because the energy required for bending is considerably lower than that required to stretch the graphene sheet. Therefore, the oscillation vertical to the surface strongly affected the oscillation parallel to the surface. 
Although lateral oscillations vertical to the surface is considered using 
a chain of coupled oscillators, the vertical oscillation of the graphene cantilever remains to be solved using coupled oscillators.

Periodic boundary conditions are typically used to solve problems related to coupled oscillators, in both  classical and quantum cases \cite{Ford1965}. However, the linear increase in the variance obtained in this study has never been observed in coupled oscillators with periodic boundary conditions. Other inherent statistical properties may appear under asymmetric boundary conditions.  

\section*{Data availability statements}    
The data that support the findings of this study are available from the corresponding author upon reasonable request.

\section*{Conflict of interest}
The author has no conflicts to disclose.

\begin{flushleft}
{\bf References}
\end{flushleft}

\appendix
\section{Derivations of Eqs. (\ref{EqAppendix1}) and (\ref{EqAppendix2})}
Let $N$ and $n$ be positive integers. We express the following summation in terms of $N$ and $n$:
\begin{eqnarray}
S_{n,N} \equiv
\sum_{k=1}^{N} \frac{\sin^2\left(\frac{(2k-1)n\pi}{2N+1}\right)}
{1-\cos\left[\frac{(2k-1)\pi}{2N+1} \right]} .
\end{eqnarray}
Using the Euler's formula, the summation is rewritten by
\begin{eqnarray}
S_{n,N} &=&
\sum_{k=1}^{N}
\frac{2\zeta_{k}-\zeta_{k}^{2n+1}-\zeta_{k}^{1-2n}}
{2(1-\zeta_{k})^2}. 
\label{SnN2}
\end{eqnarray}
Here $\zeta_{k}$ is defined by 
\begin{eqnarray}
\zeta_{k} &\equiv& (-1)^{\frac{2k-1}{M}},
\end{eqnarray}
where $M$ = $2N+1$. Because $\zeta_{k}$ is a solution of 
$x^{M}+1$, the following equation holds:
\begin{eqnarray}
x^M+1 &=& \prod_{k=1}^{M}(x-\zeta_{k}).
\end{eqnarray}
By taking the logarithm of both sides, we have
\begin{eqnarray}
\log(x^M+1) &=& \sum_{k=1}^{M}\log (x-\zeta_{k}).
\end{eqnarray}
Furthermore, by taking the derivative of both sides with respect to $x$, we obtain
\begin{eqnarray}
\frac{Mx^{M-1}}{x^M+1} &=& \sum_{k=1}^{M}\frac{1}{x-\zeta_{k}}.
\label{A6}
\end{eqnarray}
For $1 \leq k \leq N$, the following relation holds
\begin{eqnarray}
\frac{1}{x-\zeta_{k}} &=&\left(\frac{1}{x-\zeta_{M-k-1}}\right)^{\ast}.
\label{A7}
\end{eqnarray}
where $z^{\ast}$ denotes complex conjugate of $z$.
Accordingly, combining Eqs. (\ref{A6}) and (\ref{A7}) yields

\begin{eqnarray}
\mbox{Re}
\left[
\sum_{k=1}^{N}\frac{1}{x-\zeta_{k}} 
\right]
&=& 
\frac{1}{2}\left(\frac{Mx^{M-1}}{x^M+1}-\frac{1}{x+1}\right). 
\label{sumtrans}
\end{eqnarray}

Let us consider the summation of geometric progression with initial value $x^{l-1}$ for $l$ $>$ 0 and common ratio $\zeta_{k}/x$,
\begin{eqnarray}
Q_{k,l}& \equiv & \sum_{j=0}^{l-1}x^{l-1}\left(\frac{\zeta_{k}}{x}\right)^j, \\
&=& \frac{x^{l}-\zeta_{k}^{l}}{x-\zeta_{k}}.
\label{sumQk2}
\end{eqnarray}
The summation of $Q_{k}$ from $k$ = 1 to $N$ is written as 
\begin{eqnarray}
\sum_{k=1}^{N}Q_{k,l}
&=& \sum_{j=0}^{n-1}x^{n-1-j}\sum_{k=1}^{N} \zeta_{k}^j.
\label{sumQ}
\end{eqnarray}
The last summation is expressed as
\begin{eqnarray}
\mbox{Re}
\left[\sum_{k=1}^{N} \zeta_{k}^j
\right] = 
\left \{
\begin{array}{ll}
N, & j = 0, \\
\frac{1}{2}, & j = \mbox{positive odd integers}, \\
-\frac{1}{2}, & j = \mbox{positive even integers}. \\
\end{array}
\right .
\end{eqnarray}
Thus, the real part of the left-hand side of Eq. (\ref{sumQ}) for odd $l$ is 
\begin{eqnarray}
\mbox{Re}
\left[
\sum_{k=1}^{N}Q_{k,l} 
\right]
&=&Nx^{l-1}-\frac{1}{2}\frac{1-x^{l-1}}{1+x}.
\label{resumQ}
\end{eqnarray}
In the following, $l$ is assumed to be an odd number.
From Eq. (\ref{sumQk2}), the real part of the summation $Q_{k}$ is
divided into two parts as follows:
\begin{eqnarray}
\mbox{Re}
\left[
\sum_{k=1}^{N}Q_{k,l} 
\right]
&=&\mbox{Re}
\left[
\sum_{k=1}^{N}
\frac{x^{l}}{x-\zeta_{k}}
\right]
-
\mbox{Re}
\left[
\sum_{k=1}^{N} 
\frac{\zeta_{k}^{l}}{x-\zeta_{k}}
\right].
\end{eqnarray}
Using Eqs. (\ref{sumtrans}) and (\ref{resumQ}), we have
\begin{eqnarray}
\mbox{Re}
\left[
\sum_{k=1}^{N} 
\frac{\zeta_{k}^{l}}{x-\zeta_{k}}
\right]
&=&\mbox{Re}
\left[
\sum_{k=1}^{N} 
\frac{x^{l}}{x-\zeta_{k}}
\right]
-
\mbox{Re}
\left[
\sum_{k=1}^{N}Q_{k} 
\right], \\
&=&
\frac{x^l}{2}\left(\frac{Mx^{M-1}}{x^M+1}-\frac{1}{x+1}\right) 
-Nx^{l-1}+\frac{1-x^{l-1}}{2(1+x)}. 
\label{Qdivide2}
\end{eqnarray}
By taking the derivative of both sides of Eq. (\ref{Qdivide2}) sides with respect to $x$ and 
substituting $x$ = 1, the following equation is obtained:
\begin{eqnarray}
\mbox{Re}
\left[
\sum_{k=1}^{N} 
\frac{\zeta_{k}^{l}}{(1-\zeta_{k})^2}
\right]
&=&
\frac{1}{4}(l-1-4N+2lN-2N^2).
\label{R1}
\end{eqnarray}
The summation with negative powers can be obtained as 
\begin{eqnarray}
\mbox{Re}
\left[
\sum_{k=1}^{N} 
\frac{\zeta_{k}^{-l}}{(1-\zeta_{k})^2}
\right]
&=&
\mbox{Re}
\left[
\sum_{k=1}^{N} 
\frac{\zeta_{k}^{l+2}}{(1-\zeta_{k})^2}
\right].
\end{eqnarray}
The power exponent of $\zeta_{k}$ in the denominator of the left-hand side 
in Eq. (\ref{SnN2}) is odd. Thus, by replacing the summation of the right-hand side 
of Eq. (\ref{SnN2}) with the right-hand side of  Eq. (\ref{R1}),
the summation is given by
\begin{eqnarray}
\sum_{k=1}^{N} \frac{\sin^2\left(\frac{(2k-1)n\pi}{2N+1}\right)}
{1-\cos\left[\frac{(2k-1)\pi}{2N+1} \right]} 
&=& \frac{n(2N+1)}{2}.
\label{main}
\end{eqnarray}
This summation is generalized as 
\begin{eqnarray}
\sum_{k=1}^{N} 
\frac{
\sin\left(\frac{(2k-1)l_{1}\pi}{2N+1}\right)
\sin\left(\frac{(2k-1)l_{2}\pi}{2N+1}\right)
}
{1-\cos\left[\frac{(2k-1)\pi}{2N+1} \right]} 
&=&\frac{\zeta_{k}^{\alpha+1}+\zeta_{k}^{-\alpha+1}
-
\zeta_{k}^{\beta+1}-\zeta_{k}^{-\beta+1}}
{2(1-\zeta_{k})^2},
\end{eqnarray}
where $N$ $\geq$ $l_{1}$ $\geq$ $l_{2}$ and 
$\alpha$ = $l_{1}+l_{2}$, $\beta$ = $l_{1}-l_{2}$.
Using Eq. (\ref{R1}), we obtain
\begin{eqnarray}
\sum_{k=1}^{N} 
\frac{
\sin\left(\frac{(2k-1)l_{1}\pi}{2N+1}\right)
\sin\left(\frac{(2k-1)l_{2}\pi}{2N+1}\right)
}
{1-\cos\left[\frac{(2k-1)\pi}{2N+1} \right]} 
&=& \frac{l_{2}(2N+1)}{2}.
\end{eqnarray}
As a special case, Eq. (\ref{main}) is obtained by setting $l_{1}$ = $l_{2}$ = $n$.


\begin{thebibliography}{99}
\bibitem{Mazur1960}
Mazur P and Montroll E
1960
Poinca\'e cycles, ergodicity, and irreversibility in assemblies of coupled harmonic oscillators
{\it J. Math. Phys.} 
{\bf 1}
70 


\bibitem{Poggi1997}
Poggi P and Ruffo S
1997
Exact solutions in the FPU oscillator chain
{\it Physics D}
{\bf 103}
251

\bibitem{Berman2005}
Berman G P and Izrailev F M
2005
The Fermi-Pasta-Ulam problem: Fifty years of progress
{\it Chaos} 
{\bf 15} 
015104 

\bibitem{Baldovin2023}
Baldovin M, Marino R and Vulpiani A
2023
Ergodic observables in non-ergodic systems: The example of the harmonic chain
{\it Physica A}
{\bf 630}
129273


\bibitem{Takizawa1964}
Takizawa E and Kobayashi K
1964
Heat flow in a system of coupled harmonic oscillators
{\it Prog. Th. Phys.} 
{\bf 31}
1176

\bibitem{Lepri2003}
Lepri S, Livi R and Politi A
2003
Thermal conduction in classical low-dimensional lattices
{\it Phys. Rep.}
{\bf 377}
1

\bibitem{Dhar2008}
Dhar A
2008
Heat transport in low-dimensional systems
{\it Adv. Phys.}
{\bf 57}
457

\bibitem{Novotny2010}
Novotny L
2010
Strong coupling, energy splitting, and level crossings: A classical perspective
{\it Am. J. Phys.} 
{\bf 78} 
1199

\bibitem{Heagy1994}
Heagy J F, Carroll T L and Pecora L M
1994
Synchronous chaos in coupled oscillator systems
{\it Phys. Rev. E} 
{\bf 50} 
1874


\bibitem{Shim2007}
Shim S B, Imboden M and Mohanty P
2007
Synchronized oscillation in coupled nanomechanical oscillators
{\it Science}
{\bf 316}
95



\bibitem{Wang1945}
Wang M C and Uhlenbeck G E
1945
On the theory of the Brownian motion II
{\it Rev. Mod. Phys.}
{\bf 17}
323

\bibitem{Spohn1980}
Spohn H
1980
Kinetic equations from Hamiltonian dynamics: Markovian limits
{\it Rev. Mod. Phys.} 
{\bf 52}
569


\bibitem{Ford2001}
Ford G W and O'Connell R F
2001
Exact solution of the Hu-Paz-Zhang master equation
{\it Phys. Rev. D}
{\bf 64}
105020


\bibitem{Kupfermana2004}
Kupfermana R and Stuartb A M
2004
Fitting SDE models to nonlinear Kac-Zwanzig heat bath models
{\it Physica D} 
{\bf 199}
279

\bibitem{Ghosh2023}
Ghosh A and Bandyopadhyay B
2023
Quantum dissipation and the virial theorem
{\it Physica A}
{\bf 625}
128999



\bibitem{Ford1965}
Ford G W, Kac M and Mazur P
1965
Statistical mechanics of assemblies of coupled oscillators
{\it J. Math. Phys.} 
{\bf 6} 
504

\bibitem{Bao2011}
Lu H and Bao J D
2011
Time evolution of a Harmonic chain with fixed boundary conditions
{\it Chin. Phys. Lett.}
{\bf 28}
040505

\bibitem{Zou2016}
Zou J H, Ye Z Q and Cao B Y
2016
Phonon thermal properties of graphene from molecular dynamics using different potentials
{\it J. Chem. Phys.} 
{\bf 145} 
134705 


\bibitem{Eftekhari2021}
Eftekhari S A, Toghraie D, Hekmatifar M and Sabetvand R
2021
Mechanical and thermal stability of armchair and zig-zag carbon sheets using classical MD simulation with Tersoff potential
{\it Physica E}
{\bf 133}
114789


\bibitem{Bachtold2022}
Bachtold A, Moser J and Dykman M I
2022
Mesoscopic physics of nanomechanical systems
{\it Rev. Mod. Phys.} 
{\bf 94} 
045005





\bibitem{Fonseca2020}
Da Fonseca C M and  Kowalenko V
2020
Eigenpairs of a family of tridiagonal matrices: three decades later
{\it Acta Math. Hungar.} 
{\bf 160} 
376 

\bibitem{Gonzalez2023}
Herrera-Gonz{\'a}lez I F and M{\'e}ndez-Berm{\'u}dez J A
2023
Localization properties of harmonic chains with correlated mass and spring disorder: Analytical approach
{\it Phys. Rev. E} 
{\bf 107} 
034108

\bibitem{Tersoff1986}
Tersoff J
1986
Empirical interatomic potential for carbon, with applications to amorphous carbon
{\it Phys. Rev. Lett.}
{\bf 56}
632


\bibitem{Tersoff1989}
Tersoff J
1989
Modeling solid-state chemistry: Interatomic potentials for multicomponent systems
{\it Phys. Rev. B}
{\bf 39}
5566

\bibitem{Rajasekaran2016}
Rajasekaran G, Kumar R and Parashar A
2016
Tersoff potential with improved accuracy for simulating graphene in molecular dynamics environment
{\it Mater. Res. Express}
{\bf 3}
035011


\bibitem{Bissell2021}
Bissell J J
2021
On the ubiquity of classical harmonic oscillators and a universal equation for the natural frequency of a perturbed system
{\it Am. J. Phys.} 
{\bf 89} 
1094


\bibitem{Lee2008}
Lee C, Wei X, Kysar J W and Hone J
2008
Measurement of the elastic properties and intrinsic strength of monolayer graphene
{\it Science}
{\bf 321}
385

\bibitem{Jiang2009}
Jiang J W, Wang J S and Li B
2009
Young's modulus of graphene: A molecular dynamics study
{\it Phys. Rev. B} 
{\bf 80}
113405

\bibitem{Chen2013}
Chen C and Hone J
2013
Graphene nanoelectromechanical systems
{\it Proc.  IEEE}
{\bf 101}
1766

\bibitem{Lu2009}
Lu Q, Arroyo M and Huang R
2009
Elastic bending modulus of monolayer graphene
{\it J. Phys. D}
{\bf 42}
102002

\bibitem{Wei2013}
Wei Y, Wang B, Wu J, Yang R and Dunn M L
2013
Bending rigidity and Gaussian bending stiffness of single-layered graphene
{\it Nano Lett.}
{\bf 13}
26

\bibitem{Yang2020}
Yang F, Wang M, Zhang D, Yang J, Zheng M and Li Y
2020
Chirality pure carbon nanotubes: growth, sorting, and characterization
{\it Chem. Rev.}
{\bf 120}
2694
\end{thebibliography}
\end{document}